\documentclass[aps,pra,twocolumn,floatfix,nofootinbib]{revtex4-2}
\usepackage{amsmath}
\usepackage{graphicx}
\usepackage{amssymb}
\usepackage{amsmath}
\usepackage{hyperref}
\usepackage{xcolor}

\newcommand{\half}{\ensuremath{{\textstyle\frac{1}{2}}}}
\newcommand{\rmi}{{\rm i}}

\newcommand{\rme}{{\rm e}}

\begin{document}
\title{Phase-only control of GRAPE shaped pulses}
\author{Andrew J. Baldwin}\email{andrew.baldwin@chem.ox.ac.uk}
\affiliation{Chemistry Research Laboratory, University of Oxford, Mansfield Road, Oxford OX1 3TA, UK}
\affiliation{Kavli Institute for Nanoscience Discovery, University of Oxford, Oxford OX1 3QU, UK}
\author{Jonathan A. Jones}\email{jonathan.jones@physics.ox.ac.uk}
\affiliation{Clarendon Laboratory, University of Oxford, Parks Road, Oxford OX1 3PU, UK}
\date{\today}

\begin{abstract}
We compare phase-only control with phase-and-amplitude control when designing shaped pulses using the GRAPE algorithm, and explain why phase-only control has significant advantages. Trotterisation can be used to simulate amplitude modulation with phase modulation, indicating that the two approaches are fundamentally equivalent. Pulses designed by either method can be interconverted, but phase-only optimization is faster and simpler. The resulting pulses can then be converted into phase-and-amplitude form for more robust implementation if desired.
\end{abstract}

\maketitle

Shaped pulses~\cite{Warren1988} enable better excitation profiles and more precise qubit control than simple rectangular pulses. They have found widespread application in Nuclear Magnetic Resonance (NMR) spectroscopy \cite{Freeman1998}, and more recently in quantum information processing (QIP) with NMR \cite{Jones1998c}, trapped ions \cite{Steffen2007}, and atoms \cite{Ma2020}. Early designs, such as Gaussian and Hermite pulses~\cite{Warren1984}, used simple shapes which were scaled in amplitude and time for a particular application. Later designs used numerical optimization of a carefully chosen ansatz described by a small number of parameters~\cite{Geen1991}, as direct optimization of a general pulse shape was computationally infeasible at the time. This changed with the development of the GRAPE algorithm~\cite{Khaneja2005}, an efficient method for estimating gradients, permitting much faster optimization. In combination with the rapid increase in computational power this means that direct optimization of finely digitized arbitrary pulse shapes is now practical~\cite{Joseph2023}.

These techniques were largely developed within NMR, and will be described in that context, but the interaction of a spin-\half\ nucleus with an RF field can be mapped onto many other quantum technologies, and the conclusions apply much more widely. Our key assumption is that the control Hamiltonian can be described in some frame as lying in the $xy$-plane, while the background (drift) Hamiltonian $\mathcal{H}_0$ has cylindrical symmetry around the $z$-axis. This requirement is met not only by magnetic resonance systems, but also atoms and ions, and many other possible QIP implementations.

Most authors in this area have used phase-and-amplitude control, in which both the strength of the control field and its orientation in the $xy$-plane are varied. Although constant-amplitude phase-only control has also been used to great effect in NMR \cite{Skinner2006, Manu2015, Goodwin2020, Violaris2021, Haller2022, Subrahmanian2022, Joseph2023, Buchanan2025, DiPietra2025, Lim2026}, neutral atoms  \cite{Saywell2018, Saywell2020, Jandura2022, Jandura2023, Evered2023}, and ions \cite{Milne2020, Liu2023}, there is a widespread belief that amplitude and phase provide distinct control routes, and that using all the control parameters available must offer more flexibility than more restricted choices. Here we explain why phase-only control has essentially no loss in controllability, and also leads to faster and more robust optimization. 

We draw a distinction here between shaped pulses, which permit selective control of individual qubits, and composite pulses, which aim to achieve robust control in the presence of experimental imperfections, although this separation is not always entirely clear. Traditional composite pulses~\cite{Levitt1986} contain a small number of sub-pulses, each of which has a large rotation angle, frequently $90^\circ$ or $180^\circ$. By contrast shaped pulses have a much larger number of sub-pulses with much smaller rotation angles. This distinction is blurred by some more recent composite pulse designs which aim to achieve robustness to time-dependent noise~\cite{Fauseweh2012, Ball2015}. Although these pulses were originally described as frequency modulated, they are in practice implemented as phase modulated pulses. However the aim of robust control, rather than selective control, still distinguishes them from the shaped pulses considered here. Related ideas have also been explored in the context of meridional composite pulses~\cite{Bodenstedt2022}.

First we outline the basic method of GRAPE optimization, and explain why phase-only control leads to faster and better optimization. Secondly we show how amplitude modulation can be simulated by phase modulation, proving that phase-only control is always sufficient. Thirdly we describe a simple example of a phase-only pulse for use in conventional NMR. Fourthly we describe how phase-only control leads to a simpler control landscape by removing unnecessary, and therefore potentially confusing, search directions. Finally we show how concerns about the rapid phase variations and high power control fields used in phase-only pulses can be overcome by converting phase-only pulses into phase-and-amplitude form.

\section{GRAPE}
The control Hamiltonian can be taken to be piecewise constant with a time step $\tau$, and described as a linear sum over some underlying controls, with amplitude $\beta^k_j$ for the $k^{\rm th}$ control in time step $j$. The evolution during a single time step is given by 
\begin{equation}
V_j=\exp(-\rmi\mathcal{H}_j\tau)\quad\text{with}\quad
\mathcal{H}_j=\mathcal{H}_0+\sum_k\beta^k_j\mathcal{H}_k,
\end{equation}
and the total evolution is the time-ordered product 
\begin{equation}
V=V_n \dots V_{j+1} V_j V_{j-1} \dots V_1
\end{equation}
over all $n$ time steps. For unitary control in QIP we seek to minimize an infidelity
\begin{equation}
\mathcal{I}=1-|\text{Tr}(U^\dag V)/\text{Tr}(U^\dag U)|^2\label{eq:Infid}
\end{equation}
between $V$ and the target propagator $U$. Because only $V_j$ depends directly on $\beta^k_j$ it follows that
\begin{equation}
\frac{\partial V}{\partial\beta^k_j}=\left(V_n \dots V_{j+1}\right) \frac{\partial V_j}{\partial\beta^k_j} \left(V_{j-1} \dots V_1\right)
\end{equation}
and the problem reduces to calculating the gradients of the individual sub-propagators. The partial products can be calculated efficiently at the same time as $V$, and then stored for reuse, enabling gradients to be calculated in a time linear in the number of control points rather than the quadratic dependence observed for naive methods~\cite{Khaneja2005}. A similar approach can also be used to design state-to-state pulses and to optimize transfers in the presence of relaxation by using an appropriate fidelity definition~\cite{Khaneja2005}, but these generalizations do not greatly alter the control issues considered here. 

The method can be extended to tolerate systematic uncertainty in control fields by averaging the infidelity over a range of parameters. In the same way pulses for conventional NMR can be designed to perform different operations over different offset frequency bands by averaging appropriate infidelities at a range of frequencies. For conventional NMR it is often preferable to use a simpler fidelity definition \cite{Kobzar2012}, built around $\text{Tr}(U^\dag V)$ rather than the square modulus form used in Eq.~\ref{eq:Infid}, as the global phase differences between target and control propagators are confined to factors of $\pm1$. For QIP applications there are frequently global phase differences between theoretical descriptions of propagators and their experimental implementations, and so more care is needed. An alternative approach in these cases is to use the real part of the simple fidelity \cite{SchulteHerbrueggen2011,Machnes2011}.

\subsection{Phase-and-amplitude control}
In the original paper the control Hamiltonians were angular momentum operators $F_x=\sum I_x$ and $F_y=\sum I_y$, where the sums run over all the spins which interact with the same RF field, with corresponding amplitudes $\beta^x_j$ and $\beta^y_j$. Sub-propagator gradients were estimated using 
\begin{equation}
\partial V_j/\partial\beta^k_j\approx-\rmi\tau\mathcal{H}_k V_j,
\end{equation}
but this linear form is only approximate~\cite{Jones2024}, while accurate gradient calculations would permit more sophisticated quasi-Newton optimization routines such as BFGS to be used~\cite{DeFouquieres2011}. However, calculating accurate derivatives is complicated in the general case, although analytic approaches can be used for single-spin calculations~\cite{Buchanan2025}. 

A closely related alternative is to specify the control Hamiltonian in terms of the amplitude $\alpha_j$ and phase $\phi_j$ of the RF field, reflecting how shaped pulses are usually implemented. Fundamentally this changes nothing, as $\beta^x_j=\alpha\cos\phi_j$ and so on, although care needs to be taken to handle the very different sizes of amplitude and phase, but this form is more natural when considering the use of amplitude penalties. Conventional optimization of phase and amplitude frequently leads to the amplitude rising to physically impossible levels, reflecting the time efficiency of `bang--bang' control \cite{Morton2006a,Bhole2016}, which combines instantaneous (infinite amplitude) rotations with delays, similar to `jump and return' sequences \cite{Plateau1982}. To prevent this happening it is usually necessary to penalize excessive amplitudes, but this frequently hampers effective convergence~\cite{Rowland2012}.

\subsection{Phase-only control}
A more radical alternative is to keep the overall amplitude fixed at some constant value $\alpha$, and vary only the $\phi_j$ values, that is \textit{phase-only control}~\cite{Skinner2006}. This has several major advantages over the traditional approach. The Hamiltonian is now 
\begin{equation}
\mathcal{H}_j=\mathcal{H}_0+\alpha(F_x\cos\phi_j+F_y\sin\phi_j),
\end{equation}
and the sub-propagators are most conveniently evaluated in a phase-shifted frame, using
\begin{equation}
V_j=\rme^{-\rmi\phi F_z}\rme^{-\rmi\mathcal{H}'\tau}\rme^{\rmi\phi F_z}
\quad\text{with}\quad
\mathcal{H'}=\mathcal{H}_0+\alpha F_x.
\label{eq:VGRAWME}
\end{equation}
This enables almost all the large number of matrix exponentials required for amplitude and phase control to be avoided. As $\mathcal{H}'$ is fixed the central exponential only has to be calculated once, at the start of the optimization, while the phase shift terms are diagonal in the computational basis, and so the corresponding propagators can be efficiently calculated and applied~\cite{Bhole2018}. 

The form of $\mathcal{H}'$ used here is valid as long as $\mathcal{H}_0$ commutes with $F_z$, so that it is unaffected by the phase rotations. Equivalently, the drift Hamiltonian must have axial symmetry around the $z$-axis, perpendicular to the $xy$-plane containing the control Hamiltonians. This is true within NMR where $\mathcal{H}_0$ contains Zeeman terms and spin--spin couplings, as these terms are all axially symmetric around $z$. Note that this is not restricted to the weak ($I^1_zI^2_z$) couplings seen in heteronuclear systems, but also works for strong ($I^1\!\cdot\!I^2$) homonuclear couplings as well as dipole--dipole couplings. Equivalent interactions are found in many other experimental systems, such as trapped atoms.

Even more importantly the use of Eq.~\ref{eq:VGRAWME} means that the exact sub-propagator derivatives take a particularly simple form, as 
\begin{equation}
\partial V_j/\partial\phi_j=-\rmi F_z V_j + V_j\rmi F_z=\rmi[V_j,F_z]
\end{equation}
which is trivial to calculate~\cite{Bhole2018}. This `phase trick' is not confined to propagators in Hilbert space, and an entirely equivalent expression applies to super-operator descriptions in Liouville space. Access to exact gradients enables the use of quasi-Newton methods, and so phase-only optimization will be far more efficient, requiring fewer iterations which can be calculated more rapidly.  It is also possible to calculate the Hessian in a similar way, permitting direct Newton--Raphson optimization to be used~\cite{Goodwin2016}, although we do not use that here. 

A third important advantage of phase-only control is that it completely avoids the use of penalty functions to prevent the amplitude heading to infinity, as the amplitude is simply fixed at some chosen value. Note that it is not necessary to use the same fixed value for every time step, and so it is possible to use phase-only optimization on top of some specified pattern of amplitudes~\cite{Jones2024}. While it is not immediately obvious where such a pattern might be obtained, we will see a case in Section~\ref{sec:reverse} where this is a sensible approach.

Given all these advantages it might seem surprising that phase-only control is not used more widely. This seems to reflect an intuition that phase-and-amplitude control is significantly more flexible, and a consequent concern that phase-only control might not find good solutions. Next we explain why phase-only control is sufficient, giving confidence in its widespread application.

\section{Simulated amplitude modulation}\label{sec:SAM}
Here we show how amplitude modulation can be accurately simulated by phase modulation at fixed amplitude. This means that for any response which can be achieved by phase-and-amplitude control there will be a corresponding phase-only pulse sequence with very similar effects. This is not, of course, how phase-only pulses are in fact designed, but the existence of this brute-force simulation approach proves the existence of  phase-only equivalents for \textit{any} phase-and-amplitude pulse.

\subsection{Trotterisation}
For simplicity we begin with the case of on-resonance excitation of an isolated spin, for which each sub-propagator is a rotation around some axis in the $xy$-plane through some fixed angle $\theta=\alpha\tau$. For shaped pulse design we will principally be interested in excitation off-resonance, or in the presence of spin--spin interactions, or both, but it is useful to begin with the simplest case. Clearly it will not be possible to achieve much control in the case $\theta=\pi$, but for small angles it seems plausible that a much wider range of evolutions can be achieved by phase changes. This intuition can be sharpened by considering Trotter approximations~\cite{Suzuki1986}. In particular if $\mathcal{A}$ and $\mathcal{B}$ are both sufficiently small then
\begin{equation}
\rme^{\mathcal{A}}\rme^{\mathcal{B}}\approx\rme^{\mathcal{B}}\rme^{\mathcal{A}}\approx\rme^{\mathcal{A}+\mathcal{B}},\label{eq:TrotterOp}
\end{equation}
reflecting the fact that all small evolutions \textit{almost} commute. Thus for constant amplitude elements,
\begin{equation}
\theta_\phi=\rme^{-\rmi \alpha \tau I_\phi}\quad\textrm{with}\quad I_\phi=I_x\cos\phi+I_y\sin\phi,
\end{equation}
a pair of elements $\theta_{+\phi}\,\theta_{-\phi}$ can be approximated by
\begin{equation}\label{eq:Trotter}
\rme^{-\rmi \alpha \tau I_{+\!\phi}}\rme^{-\rmi \alpha \tau I_{-\!\phi}}\approx\rme^{-\rmi \alpha \tau (I_{+\!\phi}+I_{-\!\phi})}\approx\rme^{-\rmi \alpha\cos\phi\,2 \tau\, I_x},
\end{equation}
as the $I_x$ components in $I_{\pm\phi}$ add together while the $I_y$ components cancel. To identify the error term we can used the Baker--Campbell--Hausdorff expansion of Eq.~\ref{eq:TrotterOp} to get
\begin{equation}
\begin{split}
\log(\rme^\mathcal{A} \rme^\mathcal{B})&=\mathcal{A}+\mathcal{B}+\frac{1}{2}[\mathcal{A},\mathcal{B}]\\
&+\frac{1}{12}[\mathcal{A},[\mathcal{A},\mathcal{B}]]+\frac{1}{12}[\mathcal{B},[\mathcal{B},\mathcal{A}]]+\cdots
\end{split}
\end{equation}
and so the lowest order error term 
\begin{equation}
\half[\mathcal{A},\mathcal{B}]=\rmi\sin(\phi)\cos(\phi)\alpha^2\tau^2I_z
\end{equation}
is second order in $\alpha\tau$, as expected for the Trotter form.

Thus $\theta_{+\phi}\,\theta_{-\phi}$ can be approximated by a single pulse lasting time $2\tau$ with amplitude $\alpha\cos\phi$, as long as $\tau$ is short enough that $\theta=\alpha\tau\ll1$. This generalises to any pair of phase angles, with the overall phase being the average of $\phi_1$ and $\phi_2$ and the amplitude being reduced by the cosine of half the phase difference. The process is most simply thought of in terms of averaging complex amplitudes $\alpha\rme^{\rmi\phi}$, although the equivalence is only approximate because the terms in the propagator only almost commute.

Reversing this argument, any pulse whose amplitude is some fraction $\epsilon$ of the fixed amplitude $\alpha$ can be approximated by two sub-pulses of half the pulse length with the full amplitude $\alpha$ as long as their phases are offset from the original phase by $\pm\delta=\arccos\epsilon$. That is
\begin{equation}
(\epsilon \alpha\tau)_\phi=(\alpha\tau/2)_{\phi-\delta}(\alpha\tau/2)_{\phi+\delta}+O((\alpha\tau)^2),\label{eq:Trotter2}
\end{equation}
where the two pulses can be applied in either order. The error term depends on $\sin(2\delta)$, and so the Trotter form is exact in the cases $\delta=0$ ($\epsilon=1$) and $\delta=\pi/2$ ($\epsilon=0$), where the commutators are all zero, with the error reaching a maximum at $\epsilon=1/\sqrt2\approx0.7$. Thus for on-resonance excitation any shaped pulse can be approximated by a phase-only pulse as long as the step size is halved. Jumping the phase very rapidly to simulate a low effective amplitude can also be thought of as phase-ramping the pulse away from the resonance frequency of the spins being addressed, so that it has very little effect.

The accuracy can be further improved by dividing the pulse more finely, but with four sub-pulses it matters exactly how these are arranged. Here the Trotter--Suzuki form, also called Strang splitting,
\begin{equation}
(\epsilon \alpha\tau)_\phi\approx(\alpha\tau/4)_{\phi-\delta}(\alpha\tau/4)_{\phi+\delta}(\alpha\tau/4)_{\phi+\delta}(\alpha\tau/4)_{\phi-\delta},\label{eq:TS}
\end{equation}
with the middle two sub-pulses identical, is optimal. This is still only an approximation, but the symmetric form causes the first order error from the first two pulses to cancel with the first order error from the last two pulses The surviving error terms are $O((\alpha\tau)^3)$, so this is now a very good approximation for small $\alpha\tau$.

\begin{figure}
\includegraphics{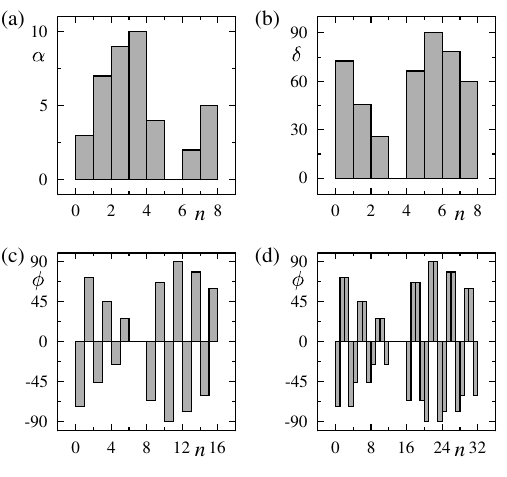}
\caption{An example of simulating an amplitude modulated pulse with phase modulation at fixed amplitude. Panel (a) shows an 8 step pulse with purely amplitude modulation, so that all sub-pulses have phase zero. The corresponding offset phases, $\delta=\arccos(\alpha/\alpha_{\rm max})$, are shown in (b) plotted in degrees: note that large amplitudes correspond to small values of $\delta$, while small amplitudes correspond to $\delta\approx90^\circ$. These offset phases can be used to build a 16 step Trotter approximation (c), or a 32 step Trotter--Suzuki approximation (d), both with $\alpha$ fixed at 10.}\label{fig:Trotter}
\end{figure}

As this process can be applied to each sub-pulse in a shaped pulse it can be used to turn an amplitude modulated pulse into a Trotter approximate phase-only pulse with twice as many sub-pulses, or a Trotter--Suzuki approximate phase-only pulse with four times as many sub-pulses, as shown in Fig.~\ref{fig:Trotter}. The fidelity of these approximations depends on $\alpha\tau$, but if $\tau$ is chosen so that the largest sub-pulse rotation angle is $10^\circ$ then the Trotter approximation (c) has an infidelity of $6\times10^{-5}$ to the original pulse (a), while the Trotter--Suzuki approximation (d) achieves $10^{-8}$. 

Above four sub-pulses there are no simple ways to improve the fidelity, as the higher order equivalents of Eq.~\ref{eq:TS} require irrational fractional pulse lengths~\cite{Yang2022}. However improvements can be made by relaxing the phase pattern slightly. In particular the form
\begin{equation}
(\epsilon \alpha\tau)_\phi=(\alpha\tau/4)_{\phi-\delta_1}(\alpha\tau/2)_{\phi+\delta_2}(\alpha\tau/4)_{\phi-\delta_1},
\end{equation}
where $\delta_1$ and $\delta_2$ now depend on $\alpha\tau$ as well as on $\epsilon$, allows amplitude modulation to be simulated perfectly. Finding the precise values of $\delta_1$ and $\delta_2$ is not simple, but they can be written in the form
\begin{equation}
\delta_1=\zeta+\xi,\quad\delta_2=\zeta-\xi,\label{eq:GPT1}
\end{equation}
with
\begin{equation}
\zeta=\arccos\left(\frac{\sin(\epsilon\alpha\tau/4)}{\sin(\alpha\tau/4)}\right)\label{eq:GPT2}
\end{equation}
and
\begin{equation}
\xi=\arctan\left(\frac{2\sin(\zeta)\cos(\zeta)\sin^2(\alpha\tau/8)}{\cos(\alpha\tau/4)+2\sin^2(\zeta)\sin^2(\alpha\tau/8)}\right).\label{eq:GPT3}
\end{equation}
Note that in the limit $\alpha\tau\rightarrow0$ we find $\zeta\rightarrow\arccos\epsilon$ and $\xi\rightarrow 0$ as expected. All the angles lie in the first quadrant, and so there is no ambiguity in the inverse trigonometric functions.

Next we consider the effects of applying pulses at a frequency $\Omega$ away from resonance, so that the Hamiltonian for each element is now $\mathcal{H}_j=\alpha I_\phi+\Omega I_z$. For the case $\Omega<\alpha$ very little changes, as the Trotter sequence 
\begin{equation}
\rme^{-\rmi \tau (\alpha I_{+\!\phi}+\Omega I_z)}\rme^{-\rmi \tau (\alpha I_{-\!\phi}+\Omega I_z)}\approx\rme^{-\rmi 2 \tau(\alpha\cos\phi\,I_x+\Omega I_z)}
\end{equation}
works in the same way as Eq.~\ref{eq:Trotter}, with the Trotter--Suzuki sequence giving the usual improvement. For $\Omega$ larger than $\alpha$ this approach begins to break down, as $\Omega\tau$ can become large even when $\alpha\tau$ is small. For large values of $\Omega$ the effect of the pulse is approximated by a Bloch--Siegert shift in the free evolution frequency by $\alpha^2/2\Omega$, which will differ for genuine and simulated amplitude modulation. As this shift shrinks with increasing $\Omega$ the accuracy of the simulation will eventually improve again. We now explore numerically how well this transformation works in practice for real pulse designs.

\subsection{Validation}
For QIP it is usually only necessary to achieve correct performance at a small number of frequencies, corresponding to individual qubits, while in conventional NMR and in atom interferometry the aim is to produce uniform performance against some specific action across one or more frequency bands. As an example we consider a phase-only implementation of the band selective E-BURP pulse~\cite{Geen1991}, which provides a good test as it only uses phases of $0^\circ$ and $180^\circ$, and so is purely amplitude modulated, but with a signed amplitude. Note that amplitude-only control cannot normally be used in QIP, as the excitation behaviour is independent of the sign of the offset frequency, but it can be used to design symmetric band-selective pulses as shown here. 

\begin{figure}
\includegraphics{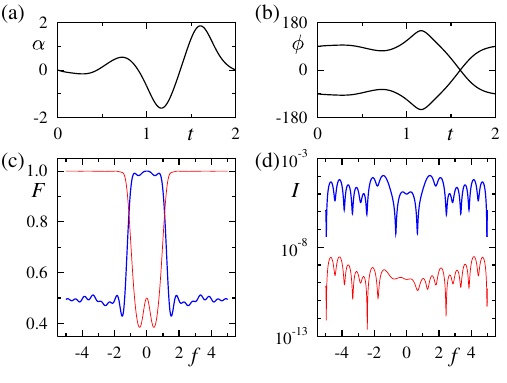}
\caption{An E-BURP pulse is shown in (a), where the signed amplitude $\alpha$ in kHz is shown as a function of time $t$ in ms. The phase envelope of the corresponding phase-only pulses (b) shows $\pm\phi$ in degrees. The actual pulse phases alternate between these values as shown in Fig.~\ref{fig:Trotter}. The state-to-state fidelity $F$ of the original pulse is shown in (c) as a function of offset frequency $f$ in kHz, with the excitation fidelity shown as a thick blue line and the miss fidelity as a narrow red line. The state-to-state fidelity between the signed amplitude original and phase-only equivalents is shown in (d) for the Trotter version (blue), and for the Trotter--Suzuki version (red).}\label{fig:EBURP}
\end{figure}

We chose a 2\,ms E-BURP pulse implemented in 1000 individual 2\,$\mu$s steps, with a maximum control amplitude of 1861\,Hz, as shown in Fig.~\ref{fig:EBURP}. This is designed to create a $90^\circ$ excitation pulse, taking $+z$ to $-y$, within a band of $\pm900$\,Hz around zero, and to avoid excitation (leaving $+z$ along $+z$) outside $\pm1500$\,Hz. The out of band `miss' propagator can be an arbitrary $z$-rotation, and as the main errors arising from the conversion to phase-only pulses are $z$-rotations, achieving high state-to-state fidelities is relatively easy.

The conventional amplitude-modulated pulse is shown in Fig.~\ref{fig:EBURP}(a), with corresponding phases shown in (b). The actual pulse phases alternate between $\pm\phi$, and so we plot the envelope within which this alternation occurs, rather than trying to show every single phase value. Note that phases outside the range $\pm90^\circ$ seen in Fig.~\ref{fig:Trotter} correspond to time points where the signed-amplitude is negative. The state-to-state fidelities of the amplitude modulated pulse are shown in (c) for the excitation fidelity (thick blue line) and miss fidelity (thin red line), revealing clearly the excitation and miss bands. As the corresponding phase-only pulses gave almost indistinguishable results we plot the state-to-state infidelity (d) between the amplitude modulated original and the phase-only equivalents. The Trotter infidelity (thick blue line) is good, while the Trotter--Suzuki infidelity (thin red line) is essentially perfect. The time step required is reduced from 2\,$\mu$s for the amplitude-modulated pulse to 1\,$\mu$s for the Trotter pulse and 500\,ns for the Trotter--Suzuki pulse. This is too short for early NMR spectrometers~\cite{Rowland2012}, but perfectly practical with modern equipment, where an underlying time step of 100\,ns or even 50\,ns is typical.

\section{Phase-only pulses}
\label{sec:POPdd}
We demonstrated above that we can convert a phase-and-amplitude modulated pulse into a phase-only pulse with almost identical performance, at the cost of having more and shorter sub-pulses than the original. The same approach will work for any shaped pulse, as the transform is entirely general. Note that the resulting infidelities are upper bounds on the errors from phase-only pulses, as they use the simple Trotter and Trotter--Suzuki forms with analytic phases. The accuracy could only be improved using numerical optimization of the individual phases of each sub-pulse, and thus designing a phase-only pulse directly. More importantly direct design can create phase-only pulses with fewer longer sub-pulses. In our own research we use phase-only control both for NMR QIP experiments and for conventional spectroscopy, and we have \textit{always} been able to find a suitable pulse. In most cases a time step of 2\,$\mu$s has proved adequate, although for pulses with very large offset frequency ranges a shorter time step is required, as expected from Nyquist arguments.

As an example we consider the direct design of a phase-only equivalent to the E-BURP pulse. For a fair comparison we use equivalent pulse parameters, that is 1000 steps each lasting 2\,$\mu$s, with a fixed control amplitude of 1861\,Hz. Our target is to create the bands identified in Fig.~\ref{fig:EBURP}, rather than to simulate an E-BURP pulse. These targets are a state-to-state $90^\circ$ excitation (from $+z$ to $-y$) for $\Omega$ within the range $\pm900$\,Hz, and no excitation ($z$ stays along $z$) from $\pm1500$\,Hz out to $\pm5000$\,Hz, with no target specified for the two transition regions between the bands. This is demanding because the width of the transition region (600\,Hz) is only just larger than the inverse of the pulse length (500\,Hz).

A pulse was designed using Seedless~\cite{Buchanan2025}, which is a very efficient implementation of single-qubit phase-only control. The infidelity was minimized over three different amplitudes, at 95\%, 100\%, and 105\% of the nominal value, simulating the effects of inhomogeneity in the control field. This can lead to error-tolerant pulses, but remains useful when, as here, the pulse is too constrained to allow full error tolerance, as it prevents a solution being found which is over-sensitive to amplitude errors. 

\begin{figure}[t]
\includegraphics{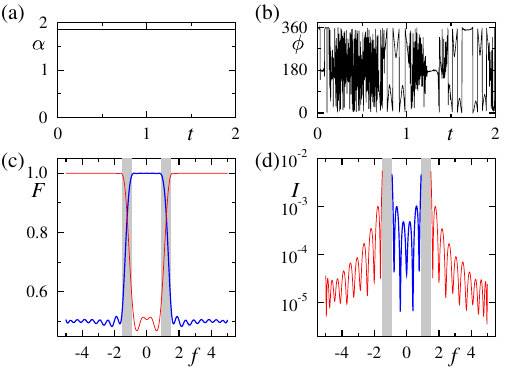}
\caption{A phase-only pulse designed to achieve similar effects to the E-BURP pulse in Fig.~\ref{fig:EBURP}. The amplitude (a) is fixed at 1861\,Hz, while the phase (b) is optimized, leading to a complicated pattern. The fidelity of this pulse is shown in (c) for excitation (blue line) and miss (red), showing the two bands in the design, with the uncontrolled transition region indicated in gray. The infidelity against an ideal pulse is shown on a logarithmic scale in (d), showing excellent behaviour except near the transition region.}\label{fig:ebpo}
\end{figure}

The results are shown in Fig.~\ref{fig:ebpo}, with the pulse shape drawn in (a) and (b). The performance at nominal amplitude is depicted as the fidelity against the two targets in (c), showing clearly the excitation and miss bands, and the transition region between them, indicated in gray. In particular the performance is better than the original E-BURP pulse, particularly at the edges of the excitation region. The infidelity against an ideal pulse, which is not defined in the transition region, is shown on a logarithmic scale in (d). The performance is excellent except in the immediate vicinity of the transition region, reflecting the challenging nature of the requirements set.

\section{Convergence}
A common complaint about optimal control theory is that convergence is unreliable, with searches becoming trapped in local minima \cite{Bondar2025}. This is perhaps surprising, as theoretical analyses indicate that quantum control should not suffer from such traps \cite{Rabitz2004,Fan2025}. There are, however, loopholes in this argument~\cite{Rach2015}. Firstly the average infidelity, used both for band selective pulses and for tolerance of implementation errors, does not correspond to a true quantum fidelity, and so the arguments need not apply. With phase-and-amplitude control the inclusion of penalty functions makes this even stronger. Penalties are not an issue for phase-only control, but it is not clear that the underlying argument applies to non-linear controls such as phase. Finally the absence of traps does not rule out the absence of saddle points or plateaus, which can act as effective traps, making optimization very tricky.

We explored the robustness of the phase-only design process by rerunning the optimization above with 1000 different choices of random seed. Of these 956 runs converged to equivalent solutions, with the remainder becoming stuck at one of several slightly sub-optimal points, with infidelities up to a factor of three higher. This is typical of our experience with phase-only pulses, which mostly converge to a global optimum but occasionally get stuck. These sub-optimal solutions differ from the ‘best’ solution by a small factor, which renders them perfectly acceptable for every-day use, but is not the best achievable. If reliably optimal performance is required, then this can be achieved using the `horse race' mode in Seedless~\cite{Buchanan2025}, which begins by performing a rough optimization with multiple starting seeds, and then selecting the leader for final full optimization. With this approach all of the 1000 runs converged to the same solution. The behaviour is different when designing phase-only pulses for QIP applications. In this case it is normally sufficient to perform a single optimization, but we do see clear signs of saddle points and plateaus in the control surface. If the infidelity is monitored during the optimization then brief moments of rapid progress punctuate long periods of near stasis. 

The superior convergence of phase-only control likely reflects the absence of penalty functions, but a more fundamental issue which arises when using phase-and-amplitude control combined with small step sizes is that there are too many different ways to achieve the same thing. If an evolution requires a decrease in the RF amplitude at some point then this can be achieved either by directly decreasing the amplitude or by increasing the phase separation between two adjacent pulses. The effects of these two changes will not be completely identical, but they will be very similar, creating significant ambiguity in the search direction. 

With phase-only control there will be less ambiguity, although if the pulse is divided too finely there will still be ambiguity over precisely how the phase modulation is implemented over a set of adjacent sub-pulses. This is clearly visible when comparing phase-only pulses generated with different random seeds. Although these pulses have the same infidelity and very similar structures, the precise pattern of phases varies with the choice of seed. This also explains a common observation, that GRAPE optimizations appear to converge better when the overall length of the shaped pulse is close to the limit for a given target. While a longer pulse might enable a lower optimal infidelity to be reached, the improvements are usually small~\cite{Skinner2006} and the extra flexibility can make finding that optimal point more challenging.

As previously observed \cite{Buchanan2025,Skinner2006,Violaris2021}, phase-only control normally leads to faster convergence and lower final infidelities than phase-and-amplitude control, and so is practically superior in almost every way.

\section{Reversing the transform}\label{sec:reverse}
\begin{figure}
\includegraphics{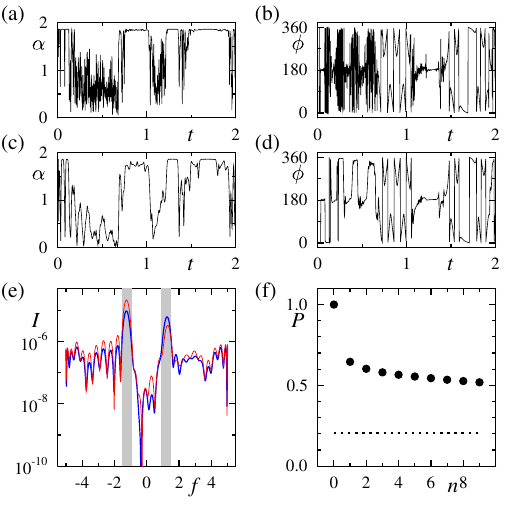}
\caption{The effect of smoothing and re-optimization of the pulse shown in Fig.~\ref{fig:ebpo}. The amplitude (a) and phase (b) are shown after one round of smoothing and re-optimization ($n=1$), and then in (c) and (d) after eight further rounds ($n=9$), showing how both amplitude and phase becomes increasingly smooth. Panel (e) shows the infidelity between the phase-only original and these smoothed versions, with $n=1$ as a thick blue line and $n=9$ as a thin red line. Note that the infidelity peaks in the gray transition regions, where the pulse design does not specify any particular result, and so deviations from the behaviour of the unsmoothed pulse are irrelevant. Finally (f) shows how the average power falls as $n$ is increased, with the power for an E-BURP pulse indicated by a dotted line.}\label{fig:ebposm}
\end{figure}
There are two final concerns which are sometimes expressed about the use of phase-only pulses, reflecting their implementation rather than their design. The first is driven by the very jagged appearance of some phase-only shaped pulses, as seen in Fig.~\ref{fig:ebpo}(b), leading to doubts as to whether control systems will be able to apply these accurately. 
In the past phase transients were an issue in NMR~\cite{Rowland2012}, leading to interest in designing smoothly varying pulses~\cite{Torosov2011}. However these issues are greatly reduced with modern control systems which use direct digital synthesis of arbitrary waveforms \cite{Momo1994,Liang2009}, and we have experienced no difficulties with step sizes around 2\,$\mu$s, and have observed acceptable behaviour with 100\,ns steps. Synthesis of amplitude modulation is also more accurate now than it used to be, but in this case concerns about amplifier linearity remain \cite{Skinner2006,Skinner2003,Skinner2004}.

The second concern is that the use of a fixed control amplitude, usually set near the maximum available, could lead to undesirable power deposition, as power varies with the square of the amplitude. While this is a genuine problem in Magnetic Resonance Imaging~\cite{Liberman2017} and some other implementations~\cite{Casanova2018}, we have found no difficulty in NMR with high-power excitations lasting tens of ms~\cite{Buchanan2025,Lim2026}. 

Both problems can be partially sidestepped by designing an initial pulse using phase-only control, and then reversing the transformation in Section~\ref{sec:SAM} to convert this to a smoother lower-power phase-and-amplitude pulse. The simplest approach is to directly average complex amplitudes of pairs of sub-pulses. Simple averaging leads to a smaller number of longer sub-pulses, but a running average can be used to create an amplitude modulated pulse with the same number of elements. As this running average is based on the Trotter form, Eq.~\ref{eq:Trotter2}, rather than the more accurate Trotter--Suzuki form, Eq.~\ref{eq:TS}, the conversion is not very accurate, but can be greatly improved by re-optimizing the phases. This requires a phase-only optimization while replacing the constant amplitude normally used with a fixed set of amplitudes obtained from the smoothed pulse. This approach is successful for a wide range of different running averages, with wider averages leading to smoother lower amplitude pulses with only a slight change in fidelity. This smoothing and re-optimization process can also be applied repeatedly, leading to a series of smoother lower-power pulses, but the power plateaus once the phases become essentially smooth.

An example is shown in Fig.~\ref{fig:ebposm} for the E-BURP style pulse discussed in Section~\ref{sec:POPdd}. The original phase-only pulse was smoothed with a 1:4:6:4:1 running average, retaining the original peak amplitude, and the phases were re-optimized. The process was then repeated eight further times, so that the number of smoothings $n$ varied from 1 to 9. Pulse amplitudes are shown in (a) and (c) with phases shown in (b) and (d) for $n=1$ and $n=9$. As shown in (e) re-optimization leads to negligible fidelity loss at each stage, while (f) shows how each smoothing leads to a reduction in integrated power. The power savings are dominated by the first smoothing, but repeated smoothing leads to pulses with much slower variations in amplitude and phase, which will be less experimentally demanding to implement. The average power does not reach the level achieved by a traditional E-BURP pulse, but this is unsurprising as the performance of the phase-only pulse is better.

\section{Conclusions}
We have shown that amplitude and phase controlled pulses can be simulated to high accuracy using phase modulation at fixed amplitude, and as a consequence it must be possible to design arbitrary shaped pulses using phase-only optimal control. This approach has many advantages, including (1) more efficient calculation of pulse propagators, speeding up the search, (2) easy calculation of exact gradients, allowing the use of efficient optimization methods such as BFGS, (3) avoiding the need for amplitude penalties which compete with the underlying optimization. The removal of unnecessary control parameters leads to a simpler control landscape, and thus much more reliable convergence. The first two speed gains are only available when the drift Hamiltonian has axial symmetry, so that Eq.~\ref{eq:VGRAWME} can be used, but the core equivalence between phase-only control and phase-and-amplitude control does not depend on this, and phase-only control remains possible even for the most general form. Concerns about implementation problems are frequently overstated, but can in any case be reduced by converting phase-only pulses to phase-and-amplitude pulses and then re-optimizing the phases. 

In our own research we have used phase-only control for all our NMR QIP experiments requiring shaped pulses since 2021. For conventional NMR spectroscopy the Seedless paper \cite{Buchanan2025} contains 54 separate pulses, covering all the main experiments in protein NMR, together with demonstrations that they perform effectively, leading to a sensitivity gain of 58\% in a HNCACO experiment. Design scripts for all of them are included in the demonstration section of the Seedless package, enabling equivalent pulses to be designed on different NMR systems. These results build on an extensive prior repertoire of phase-only pulses \cite{Skinner2006, Manu2015, Goodwin2020, Violaris2021, Haller2022, Subrahmanian2022, Joseph2023, DiPietra2025, Lim2026, Saywell2018, Saywell2020, Jandura2022, Jandura2023, Evered2023, Milne2020, Liu2023}, designed by many different methods for many different purposes. Putting all these advantages together, we believe that phase-only control should normally be the preferred approach to shaped pulse design.

\begin{acknowledgments}
We thank Mohamed Sabba and Steffen Glaser for helpful conversations. Equations \ref{eq:GPT1} to \ref{eq:GPT3} were derived with the assistance of GPT-5.6 Sol, and then checked using Mathematica. Andrew Baldwin is supported by ERC grant 101002859. For the purpose of Open Access, the author has applied a CC BY public copyright license to any Author Accepted Manuscript version arising from this submission. 
\end{acknowledgments}

\appendix

\section{Syntax for Seedless}
Phase-only pulses were produced using the Seedless GRAPE optimization package \cite{Buchanan2025} version 1.3. The pulse in Fig.~\ref{fig:ebpo} was designed using
\begin{verbatim}
frq 600. (MHz)
maxIter 3000
RF: (B1/weight)
0.95 0.25
1.00 0.5
1.05 0.25
SpinSystem: (Min/Max/Num)
Low -8.33 -2.5 50
Mid -1.5 1.5 25
High 2.5 8.33 50
Carriers: (ppm)
 0.0
wmH: (Hz)
 1861
Durations: (Total, points)
 2000e-6 1000
Targets: (Low Mid High)
 Iz Iz , Iz -Iy , Iz Iz
\end{verbatim}
The horse race mode can be implemented by including
\begin{verbatim}
Horsey 20
HorseyIter 100
\end{verbatim}
in the header. For full details see the Seedless manual.

Specifying \texttt{AMPtoPHASE} in the header will cause Seedless to output a phase only version of an amplitude modulated pulse (as a separate output), derived from the Trotter-Suzuki approximation in Eq.~\ref{eq:TS}, which increases the number of elements by 4. 

Similarly, specifying \texttt{PHASEtoAMP 5} will cause a binomial smoothing to be applied to a recently read in phase only pulse. Where the value is set to 5, the smoothing will be a normalised filter spanning 5 elements weighted 1:4:6:4:1 as described in the text. Seedless will then perform a phase only optimisation holding the amplitudes constant. The result can be successively read in and the process repeated as described in the text to create a robust replica of the original phase-only pulse in amplitude-and-phase form.

\bibliography{phaseonly.bib}
\end{document}